\documentclass[12pt]{iopart}
\usepackage{graphicx,iopams}
\usepackage[english]{babel}
\usepackage[margin=10pt,font=small,labelfont=bf]{caption}
\begin{document}
\selectlanguage{english}
\title[Magnetic non-contact friction from domain wall dynamics actuated by oscillatory...]{Magnetic non-contact friction from domain wall dynamics actuated by oscillatory mechanical motion}

\author{Ilari Rissanen$^1$ and Lasse Laurson$^{1,2}$}
\address{$^1$Helsinki Institute of Physics,
Department of Applied Physics, Aalto University, P.O.Box 11100, 
FI-00076 Aalto, Espoo, Finland.}
\address{$^2$Computational Physics Laboratory, Tampere University, P.O. Box 692, FI-33014 Tampere, Finland}
\ead{ilari.rissanen@aalto.fi}

\begin{abstract}
Magnetic friction is a form of non-contact friction arising from the dissipation of energy 
in a magnet due to spin reorientation in a magnetic field. In this paper we study magnetic 
friction in the context of micromagnetics, using our recent implementation of smooth 
spring-driven motion \mbox{[Phys. Rev. E. 97, 053301 (2018)]} to simulate ring-down measurements 
in two setups where domain wall dynamics is induced by mechanical motion. These include 
a single thin film with a domain wall in an external field and a setup mimicking a magnetic cantilever 
tip and substrate, in which the two magnets interact through dipolar interactions. We investigate how 
various micromagnetic parameters influence the domain wall dynamics actuated by the 
oscillatory spring-driven mechanical motion and the resulting damping coefficient. 
Our simulations show that the magnitude of magnetic friction can be comparable to other forms 
of non-contact friction.
For oscillation frequencies lower than those inducing excitations of the internal structure 
of the domain walls, the damping coefficient is found to be independent of frequency. Hence, our results 
obtained in the frequency range from 8 to 112 MHz are expected to be relevant also for typical experimental 
setups operating in the 100 kHz range. 
\end{abstract}

\submitto{\JPD}
\maketitle

\section{Introduction}

Friction is the ubiquitous process of mechanical energy turning into heat through various coupling mechanisms between the relative motion of objects and their internal degrees of freedom. Due to the abundance of mechanisms involved, including phononic \cite{phononic}, electronic \cite{electronic} and quantum processes \cite{quantum}, a comprehensive description of friction has proved elusive, with multiple open questions and areas of research \cite{frictionreview}. By consequence, for a relatively long period of history, the description of sliding friction was (and still is) left to simple phenomenological laws \cite{slidingkirja}. However, the advent of modern nanoscale measurement techniques such as atomic force microscopy (AFM) and friction force microscopy (FFM) have re-ignited the interest in friction and spawned the field of nanotribology, which delves into finding the fundamental constituents of frictional processes. 

Probably the most common concept associated with friction is the dry sliding friction between two solid surfaces, in which the plastic and elastic deformations of surface asperities dissipate the kinetic energy \cite{slidingkirja}. When the surfaces are separated by more than a few nanometers and can't be considered to ''touch'' each other, asperity deformation ceases to occur and the interaction of the surfaces is mediated by long range electromagnetic interactions. These interactions result in what is known as non-contact friction, a type of friction that is typically orders of magnitude weaker than contact friction, with frictional forces measured in attonewtons and damping coefficients in the \mbox{$10^{-14}$ -- $10^{-13}$ kg/s} range \cite{tribobook}. Understanding non-contact friction is not only important on the fundamental level due to the interactions being the building blocks of a comprehensive picture of friction, but also in practical sense, since the strength of non-contact friction limits the sensitivity of force sensors \cite{sensitivity}.

The electric components of electromagnetic non-contact friction, including electrostatic friction \cite{electrostatic} and Van der Waals friction \cite{vdW}, are quite well-established both experimentally and theoretically. The magnetic component, while also demonstrated experimentally \cite{atomicexperiment, tipexperiment}, has received relatively little scientific attention. In this paper, we study magnetic non-contact friction arising from dynamics of domain walls actuated by oscillatory mechanical motion via micromagnetic simulations, with a focus on the effect of the material parameters on the resulting damping coefficient.

The paper is organized as follows: In Sec. II, we briefly discuss the previous research on magnetic friction and consider the energy dissipation from a micromagnetic viewpoint. Sec. III introduces the specifics of our micromagnetic simulation setup, and Sec. IV details the results obtained from the simulations. The conclusions from this study are presented in Sec. V. 

\section{Magnetic friction}

Magnetic losses, and thus magnetic friction arise from the dissipation of energy in spin reorientation in response to a changing magnetic field inside the magnet. Though there is no general consensus on the complete nanoscale explanation for this dissipation as of yet, there are multiple pathways for the energy to be converted into heat, such as magnon-phonon and magnon-impurity interactions, and magnon scattering on the surface and interface defects \cite{dissipations}. In addition to these direct pathways of energy dissipation, other indirect magnetization-related mechanisms are also responsible for magnetic losses, such as magnetic damping (or magnetic drag), in which eddy currents inside a conductor in a changing magnetic field convert energy into heat. In this case, it is the currents and the resistivity of the conductor that are mostly responsible for the energy dissipation rather than the magnetic degrees of freedom of the magnet directly. 

The link between changing magnetization and energy dissipation on the atomic level has been demonstrated experimentally. Utilizing a spin-polarized scanning tunneling microscope, it was observed that the force required to move a magnetic adatom from an adsorption site to another increased by up to \mbox{60 \%} compared to non-magnetic adatoms \cite{atomicexperiment}. Magnetic dissipation was also shown to occur in an experiment using a soft cantilever with a magnetic tip oscillating in an external magnetic field \cite{tipexperiment}. In strong (approaching \mbox{6 T}) external magnetic fields, the damping coefficient measured for a cobalt tip were in the \mbox{$10^{-13}$ -- $10^{-12}$ kg/s} range. The damping coefficient was found to be material dependent, with magnetically more malleable cobalt showing high dissipation compared to the stronger anisotropy PrFeB, for which the damping coefficient didn't differ significantly from a bare silicon cantilever internal damping coefficient.

Magnetic friction and its dependence on parameters such as temperature have also been investigated computationally with Monte Carlo simulations in the Ising model \cite{IsingSpinFriction} and models using the Landau-Lifshitz-Gilbert equation \cite{llgFriction3, fricTemp} for the dynamics of the spins. Both velocity independent (Coulomb) friction and velocity dependent (Stokes) friction have been demonstrated, though the difference in frictional behavior could be explained by the simulation model \cite{StokesCoulomb}. Additionally, there has been a study of a larger, two-film configuration with multiple stripe domains, in which it was shown that the domain structure can evolve into a configuration that minimizes the friction \cite{llgFriction2}. As pointed out in the study, an interesting aspect of magnetic friction is that the strength of force could be adjusted by external applied fields. 

With some of the general phenomenology having been established by the aforementioned research, we focus on the magnetic losses generated by the motion of domain walls in thin films under mechanical oscillation. We employ micromagnetic simulations to observe how changes in material parameters affect the domain wall structure and the measured magnetic friction through changing the relative strength of exchange interaction and magnetic anisotropies. The simulations are performed using the micromagnetic code Mumax3 \cite{vansteenkiste2014design} with our previously developed extension \cite{extension}, in which it is possible to simulate smooth spring-driven harmonic motion of the magnet(s) 
simultaneously with the magnetization evolution.

\subsection{Micromagnetics and energy dissipation}
In the framework of micromagnetism, the behavior of the magnetic moments in a magnetic material is described by the Landau-Lifshitz-Gilbert (LLG) equation:  
\begin{equation}
\frac{\partial \textbf{m}}{\partial t}=-\frac{\gamma_0}{1+\alpha^2}\big(\mathbf{H}_{\mathrm{eff}} \times \mathbf{m} + 
\alpha \mathbf{m} \times (\textbf{m} \times \mathbf{H}_{\mathrm{eff}})\big),
\label{EQLLGNumerical}
\end{equation}
where $\mathbf{m}$ is the magnetization normalized by the saturation magnetization $M_\mathrm{sat}$ of the material (\mbox{$\mathbf{m}=\mathbf{M}/M_\mathrm{sat}$}), \mbox{$\gamma_0 = |\gamma|\mu_0 \approx 221$ kHz/(Am$^{-1}$)} is the gyromagnetic constant, where $\gamma$ is the electron gyromagnetic ratio and $\mu_0$ is the permeability of vacuum, $\alpha$ is the Gilbert damping constant, and $\mathbf{H}_{\mathrm{eff}}$ is the effective magnetic field. The effective field takes into account the exchange interaction, magnetocrystalline anisotropy, the demagnetizing field created by the magnetization itself, and the external field. The first term of the LLG equation describes the precession of a magnetic moment around the effective field, while the second term is the phenomenological relaxation term describing how the rotation of the magnetic moment eventually winds down to the direction of the effective field.

Our subject of interest is the energy loss during domain wall motion predicted by the micromagnetic theory. One can derive equation for the power dissipation with the help of the LLG equation \cite{llgFriction1}. In a micromagnetic system with the four effective field terms, the energy density of the system is
\[
\varepsilon = A_\mathrm{ex}(\nabla \mathbf{m})^2+\varepsilon_\mathrm{k}-\mu_0 M_\mathrm{sat}(\mathbf{H}_\mathrm{ext}\cdot \mathbf{m})-\frac{1}{2}\mu_0 M_\mathrm{sat}(\mathbf{H}_\mathrm{d}\cdot \mathbf{m}),
\]
where $A_\mathrm{ex}$ is the exchange constant, $\varepsilon_\mathrm{k}$ is the magnetocrystalline anisotropy energy density, $\mathbf{H}_\mathrm{ext}$ is the external field, and $\mathbf{H}_\mathrm{d}$ is the demagnetizing field \cite{numericalmicromagnetics}. 

In the absence of an external field varying in time, the only time-dependence comes from the magnetization $\mathbf{m}$ (since $\varepsilon_\mathrm{k}$ and $\mathbf{H}_\mathrm{d}$ also depend on $\mathbf{m}$). As such, the change of energy density in time can be written as
\[
\frac{d\varepsilon}{dt} = \frac{\delta \varepsilon}{\delta \mathbf{m}}\frac{\partial \mathbf{m}}{\partial t} =  -\mu_0 M_\mathrm{sat} \mathbf{H}_\mathrm{eff}\cdot\frac{\partial \mathbf{m}}{\partial t},
\]
where the definition of the effective field as the functional derivative of the energy density with respect to the magnetization \cite{numericalmicromagnetics} is used for the latter part. Inserting the LLG equation from (\ref{EQLLGNumerical}) in place of the derivative $\partial \mathbf{m}/\partial t$, we obtain
\[
\frac{d\varepsilon}{dt} = \frac{\mu_0 \gamma_0  M_\mathrm{sat}}{1+\alpha^2}\mathbf{H}_\mathrm{eff}\cdot \big(\mathbf{H}_{\mathrm{eff}} \times \mathbf{m} + 
\alpha \mathbf{m} \times (\textbf{m} \times \mathbf{H}_{\mathrm{eff}})\big),
\]
which, noting that the first term is zero and using some vector calculus identities, can be simplified to
\[
\frac{d\varepsilon}{dt} = \frac{\alpha\mu_0 \gamma_0 M_\mathrm{sat}}{1+\alpha^2}(\mathbf{m} \times \mathbf{H}_{\mathrm{eff}})^2
\]
The dissipated energy in a unit of time due to spin relaxation, or ''Gilbert dissipation'' \mbox{power $P$} inside the volume $V$ is then  
\begin{equation}
P = \frac{dE}{dt} = \int_V \frac{d\varepsilon}{dt}~dr^3= \int_V \frac{\alpha\mu_0 \gamma_0 M_\mathrm{sat}}{1+\alpha^2}(\mathbf{m} \times \mathbf{H}_{\mathrm{eff}})^2~dr^3.
\end{equation}
This expression is equivalent to the expression formulated by Brown \cite{brown1963micromagnetics}, 
\begin{equation}
P = \frac{\alpha \mu_0 M_\mathrm{sat}}{\gamma_0}\int_V\Big(\frac{d\mathbf{m}}{dt}\Big)^2 dr^3,
\label{EQBrown}
\end{equation}
only defined through instantaneous quantities $\mathbf{m}$ and $\mathbf{H}_\mathrm{eff}$ rather than the time derivative of the magnetization.

In finite difference micromagnetics, the magnet is typically discretized into cells of equal volume, \mbox{with $\mathbf{m}$} and $\mathbf{H}_\mathrm{eff}$ being constant in each cell. In such a system, the total dissipation power is then just a sum over the dissipation in individual cells:
\begin{equation}
P = \frac{ \alpha\mu_0 \gamma_0 M_\mathrm{sat} V_\mathrm{cell}}{1+\alpha^2}\sum_{i=1}^N\big(\mathbf{m}_\mathrm{i}\times \mathbf{H}_\mathrm{eff,i}\big)^2,
\label{EQDissipation}
\end{equation}
where $V_\mathrm{cell}$ is the volume of the discretization cell. From \mbox{(\ref{EQDissipation})} one can see that the material dependent \mbox{parameters $\alpha$} \mbox{and $M_\mathrm{sat}$} are explicitly present, influencing the observed energy dissipation and, by extent, the magnetic friction. Additionally, the exchange \mbox{constant $A_\mathrm{ex}$} and the uniaxial anisotropy \mbox{constant $K_\mathrm{u}$} are important parameters in our simulations, since they determine the width of the domain walls, which is of the order $\sqrt{A_\mathrm{ex}/K_\mathrm{u}}$ \cite{magneticmicrostructures}. However, these factors are hidden in the effective field term in \mbox{(\ref{EQDissipation})}, and determining the exact effect the various field terms have on magnetic friction is nontrivial.

In large systems of thousands or millions of interacting magnetic moments, the complex time evolution of the effective field and magnetization make it unfeasible to study magnetic friction analytically. Thus we approach the problem by simulating ring-down measurements, in which damping is measured through the gradual diminishing of the amplitude of mechanical oscillations of e.g. a cantilever under the presence of damping effects. Experimental ring-down measurements have been performed to measure various forms of non-contact friction, including electric friction between gold-coated substrate and \mbox{tip \cite{electrostatic2}} and dielectric friction between polymers \cite{dielectricFriction}.

\section{Micromagnetic simulation setup}

\begin{figure}
\begin{center}
\includegraphics[trim=1.1cm 6.7cm 3.5cm 7.5cm, clip=true,width=0.5\columnwidth]{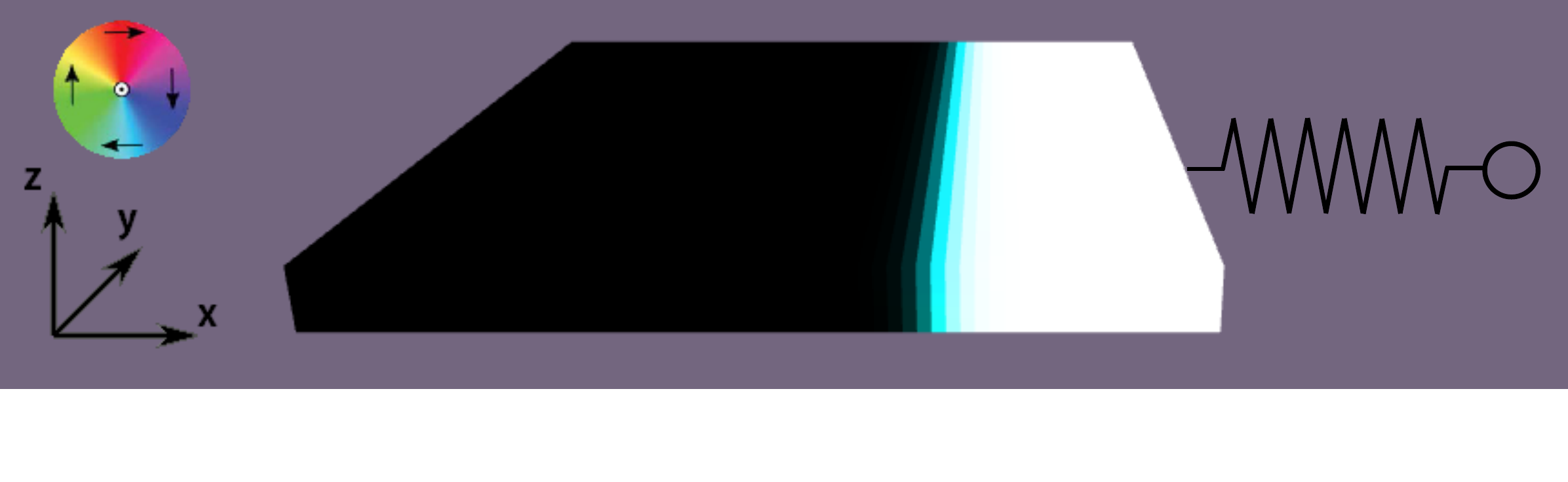}
\includegraphics[trim=0.1cm 3.68cm 0cm 2.5cm, clip=true,width=0.43\columnwidth]{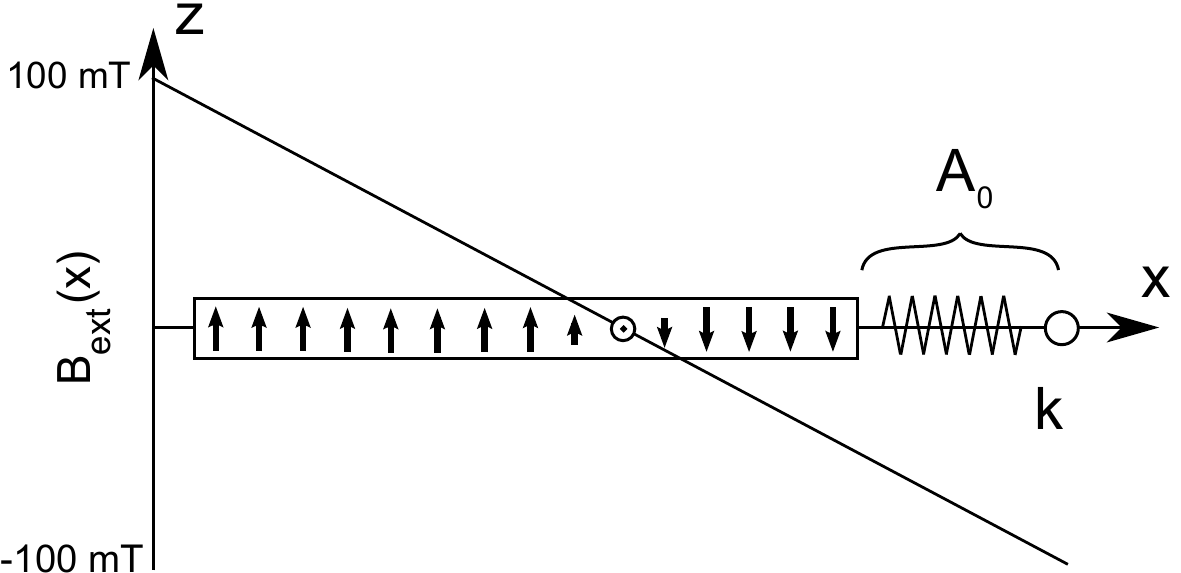}
\caption{The single film setup with a single Bloch domain wall (left). The color wheel shows the in-plane magnetization direction, whereas black and white indicate the $+z$ and $-z$-directions. A schematic image (right) shows the external field as a function of $x$. The domain wall is centered at approximately the location where $B_\mathrm{ext}(x) = 0$. \vspace{-5mm}}
\label{FIGsetup1}
\end{center}
\end{figure}

\begin{figure}
\begin{center}
\includegraphics[trim=5.3cm 7cm 5.7cm 8cm, clip=true,width=0.65\columnwidth]{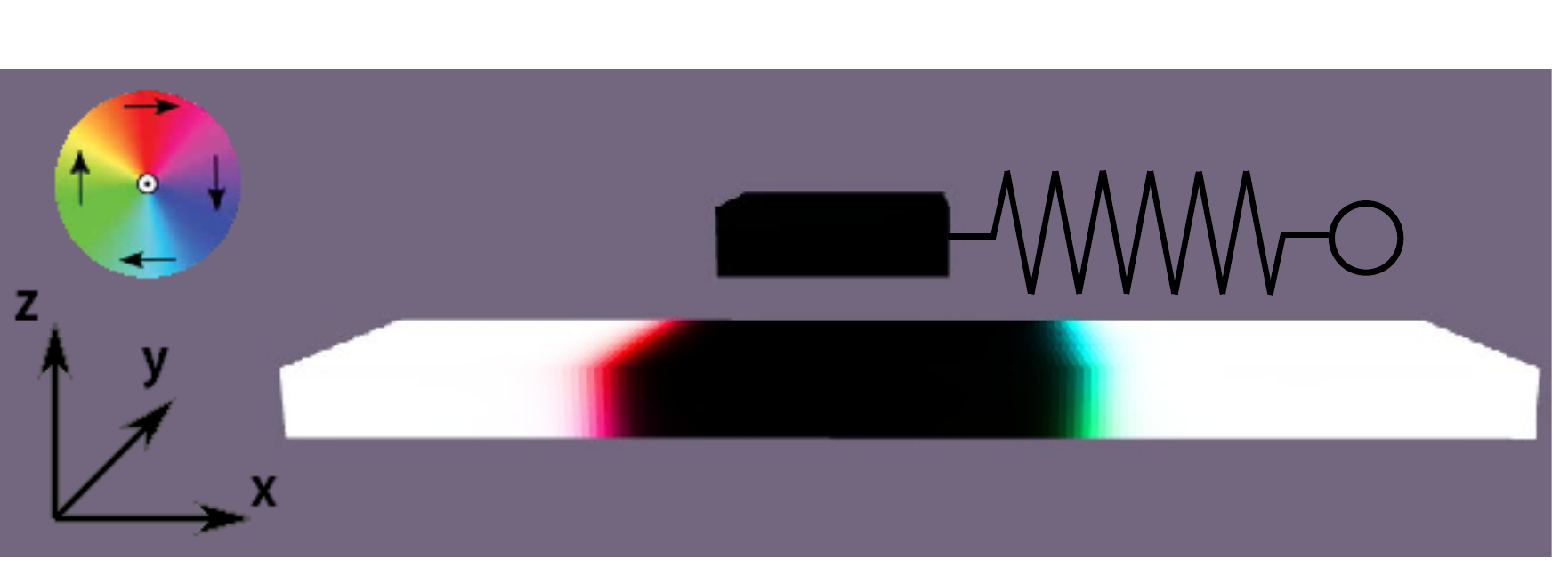}
\caption{The tip+strip setup, with \mbox{a 60 x 60 nm} square film depicting the tip of a cantilever and a thin magnetic strip with two Bloch domain walls as a base, with a 20 nm distance between the tip and the base.}
\label{FIGsetup2}
\end{center}
\end{figure}

In our ring-down simulations, we employ two simulation setups coupling domain wall dynamics with mechanical motion: 
A single thin film with a domain wall in a spatially varying external magnetic field \mbox{(figure~\ref{FIGsetup1})}, 
and a configuration mimicking a small magnetically coated tip of an oscillating cantilever and a strip of magnetic 
material containing two domain walls as a substrate \mbox{(figure~\ref{FIGsetup2})}. The film (or tip, depending on 
the setup) is attached to a spring. Like in experimental ring-down measurements, the motion of the film/tip is 
treated as damped harmonic oscillation
\begin{equation}
m\frac{d^2x}{dt^2}+\Gamma\frac{dx}{dt}+kx = F_x,
\label{EQOscillation}
\end{equation}
where $m$ is the mass of the film/tip, $\Gamma$ is the damping coefficient and $k$ is the spring constant. The 
force $F_x$ is the $x-$directional component of the magnetic \mbox{force $\mathbf{F}$} exerted on the film/tip,
\[
\mathbf{F} = \mu_0M_\mathrm{sat}V_\mathrm{cell} \sum_{i=1}^N\nabla(\mathbf{m}_\mathrm{i}\cdot\mathbf{H}_\mathrm{e,i}),
\label{EQForce}
\]
where $\mathbf{H}_\mathrm{e,i}$ is the external field in first setup and the demagnetizing field of the substrate 
in the second setup, and the sum goes over the cells of the film or tip in their respective setups. The film/tip is constrained to move only in the $x-$axis, hence only the $x-$component of the force is required.


A real oscillator has an internal friction \mbox{coefficient $\Gamma_0$}, which in the case of a cantilever can be measured by having the cantilever oscillate in a vacuum and as isolated from external influences as possible. Typical internal damping coefficients are of the \mbox{order $\Gamma_0 = 10^{-14}-10^{-13}$ kg/s} for soft cantilevers \cite{tribobook}. The total damping coefficient is then the sum of the internal damping coefficient and the damping effects of external influences $\Gamma = \Gamma_0+\Gamma_e$. An advantage of our simulations is that we can set the internal damping coefficient of the spring to zero, and thus all the energy losses come from the energy dissipation through the magnetic degrees of freedom. As such we don't explicitly simulate the $\Gamma\cdot dx/dt$ term in the simulations, but the damping term arises naturally from the magnetic dissipation instead, due to the relaxation of the magnetic moments affecting the force $F_x$ in \mbox{(\ref{EQOscillation})}.

\begin{figure}
\begin{center}
\includegraphics[trim=0cm 0cm 0cm 0cm, clip=true,width=0.80\columnwidth]{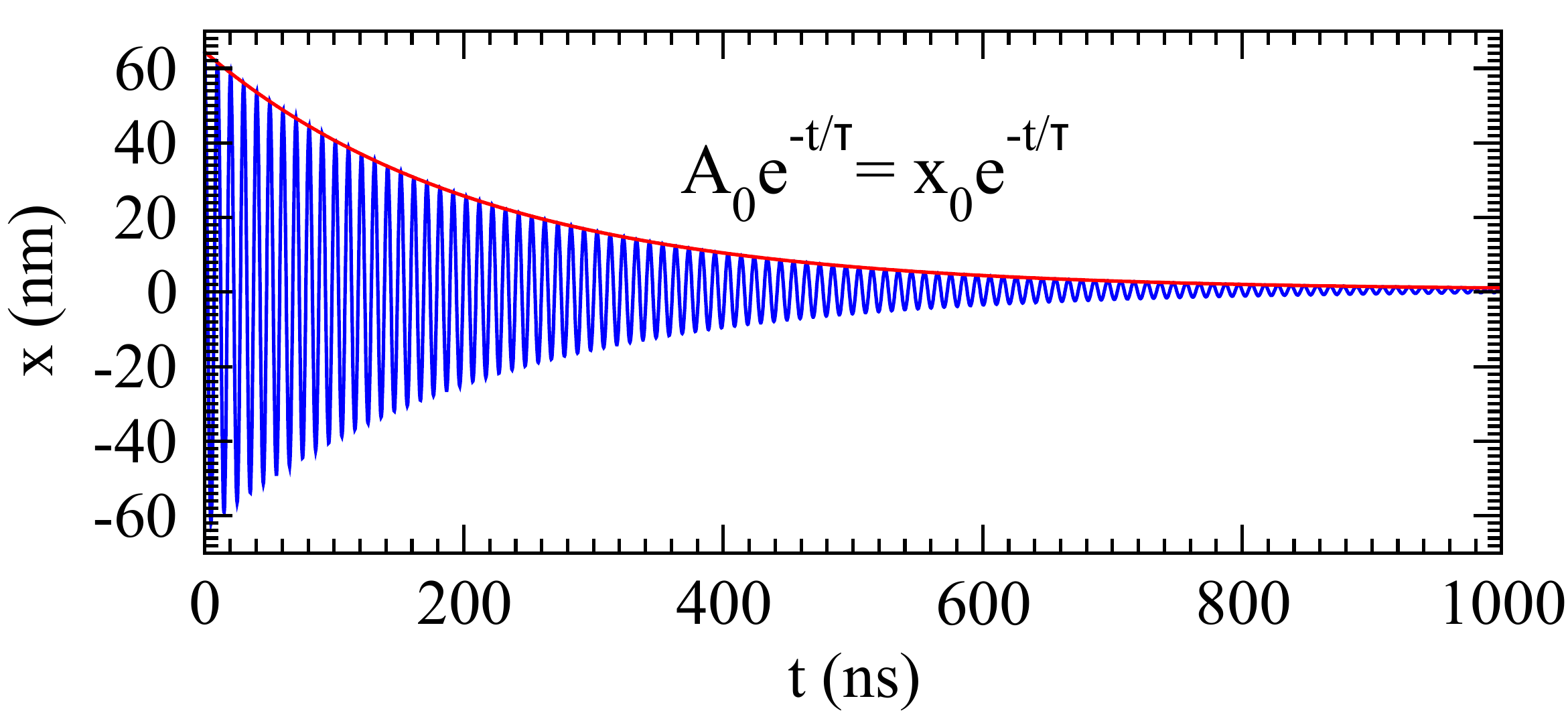}
\caption{An example of the oscillation of the position $x$ of the film/tip in a ring-down measurement with $f_0 = 112$ MHz, and an exponential function fit to the amplitude to determine the \mbox{decay time $\tau$}. The initial position of the film $x_0$ is equal to the initial amplitude of the oscillation $A_0$.}
\label{FIGringdown}
\end{center}
\end{figure}

The spring starts elongated to \mbox{length $A_0$} on the $x$-axis at the beginning of the simulation. Eventually, the oscillations die down due to the Gilbert dissipation defined \mbox{by (\ref{EQDissipation})}. At the end of a simulation, we fit an exponentially decaying function to the \mbox{amplitude $A$}
\[
A = A_0 e^{(-t/\tau)}
\]
to find the decay time $\tau$ (example in figure~\ref{FIGringdown}). The damping \mbox{coefficient $\Gamma$} is then calculated from the decay time via the relation
\[
\Gamma = \frac{2k}{\omega_0^2 \tau} = \frac{2m}{\tau},
\]
where $\omega_0$ is the natural angular oscillation frequency $\omega_0 = \sqrt{k/m}$. \cite{tribobook}

The domain wall behavior and thus magnetic friction depend on the magnetic properties of the system, which in micromagnetics are represented by material parameters. In \cite{magneticmicrostructures}, there are listed some material parameter value ranges for common magnetic materials. For both the film and the tip-strip configurations, we pick the initial material parameters from approximately the middle of these ranges, \mbox{$A_\mathrm{ex} = 5\cdot 10^{-12}$ J/m}, \mbox{$M_\mathrm{sat} =$ 350 kA/m}, \mbox{$K_\mathrm{u} = 1.2 \cdot 10^5$ J/m$^3$} and \mbox{$\alpha = 0.05$}, from here on referred to as default parameters. With these parameters, the magnetocrystalline anisotropy dominates, resulting in an out-of-plane polarized magnetization as shown in \mbox{figures~\ref{FIGsetup1} and \ref{FIGsetup2}}. We investigate how the domain wall structure and motion change when we vary each parameter at a time while keeping the rest constant. The parameters were tested in the following value ranges: \vspace{0mm}
\begin{itemize}
\item $A_\mathrm{ex}$ = $1 - 50$ pJ/m \vspace{0mm}
\item $K_\mathrm{u}$ = $0 - 300$ kJ/m$^3$ \vspace{0mm}
\item $M_\mathrm{sat}$ = $50 - 530 $ kA/m \vspace{0mm}
\item $\alpha$ = $0.001 - 0.5$. \vspace{0mm}
\end{itemize}
In micromagnetic simulations, having a discretization cell size well below magnetic exchange lengths $\sqrt{A_\mathrm{ex}/K_\mathrm{u}}$ and $\sqrt{2A_\mathrm{ex}/\mu_0 M_\mathrm{sat}^2}$ is required to not introduce numerical problems, such as artificial pinning of the domain walls, into the system \cite{numericalmicromagnetics}. In our simulations, a discretization cell size of \mbox{4 $\times$ 4 $\times$ 4 nm} sufficed for most parameters. However, for high $K_\mathrm{u}$ or low $A_\mathrm{ex}$, it's possible that the exchange length(s) shrink close to the cell size, inducing the aforementioned domain wall pinning, affecting the damping coefficient. To avoid this, for the high values of $K_\mathrm{u}$ and low values of $A_\mathrm{ex}$, we halved the cell size to \mbox{2 $\times$ 2 $\times$ 2 nm} in the single film system. We found that the tip-strip system was more sensitive to the discretization cell size, most likely due to the weaker fields involved in the domain wall movement. As such we used the smaller cell size for all simulations in that system. All simulations were run in zero temperature.

For the single thin film in an external field, we simulate a 256 $\times$ 256 $\times$ 20 nm film and an external $z$-directional field ramping linearly down from $+100$ mT to $-100$ mT. The $z$-axis was also chosen as the easy axis of the uniaxial anisotropy. With the default parameters, the resulting magnetic texture is a two-domain film with a single Bloch domain wall. We set the initial amplitude $A_0$ to 64 nm in order to have large oscillations but having the domain wall still stay relatively far from the edges of the film. 

The strip and pendulum tip geometry is more reminiscent to an actual experimental ring-down measurement setup. The cantilever tip is modeled as a square film \mbox{of 60 $\times$ 60 $\times$ 20 nm} size, while the substrate \mbox{is a 320 $\times$ 80 $\times$ 20 nm strip}. The tip and substrate are set 20 nm apart, so that we're in the non-contact friction regime and only the demagnetizing fields are relevant for the tip-substrate interaction. Multiple equilibrium magnetization configurations are possible, but we chose one with two domain walls in the strip due to it being nicely symmetric. Because of the small size of the tip and the strength of the anisotropy with the default parameters, the magnetization of the tip is forced uniformly into the anisotropy easy axis direction. The initial oscillation amplitude is set to $A_0 = 20$ nm.

To keep simulation times reasonably short, we need to have enough oscillations to dissipate energy in relatively little amount of time ($t < 1$ $\mu$s). Hence our springs in both setups have parameters outside the range of typical measurement equipment, namely a high oscillation \mbox{frequency $f_0 = \omega_0/2\pi =  56 $ MHz} 
compared to the usual \mbox{100 kHz} range of cantilevers used in typical experiments \cite{tribobook}, although very high frequency (VHF) cantilevers for high-precision force and displacement measurements can have frequencies up to tens or hundreds of MHz\cite{li2007ultra}. However, due to the rapid  magnetic relaxation, it is expected that the time scales of the mechanical motion and the magnetic relaxation are well-separated already in our simulations with most parameter configurations. Thus the change in magnetic 
structure should be in most cases independent of velocity, meaning that the observed behavior should also match lower frequency cantilevers. To see whether the frequency affects the domain wall dynamics and the damping, we also run the same simulations with a higher \mbox{frequency $f_0 =  112 $ MHz} and compare the results. Moreover, to verify the hypothesis of frequency-independent damping coefficients for low enough frequencies, we perform example simulations for the default parameters for frequencies down \mbox{to $f_0 = 8$ MHz}. 

For the single film setup, it is possible to use \mbox{(\ref{EQDissipation})} to make a prediction for the 
order of magnitude of the damping coefficient. Since in each discretization cell we have 
\mbox{$|\mathbf{m}|^2 = 1$}, and all the spin reorientation is happening inside the domain wall, 
we can write the sum of the vector products between magnetic moments and the local effective field 
in \mbox{(\ref{EQDissipation})} as \[
\sum_{i=1}^N(\mathbf{m}_\mathrm{i}\times \mathbf{H}_\mathrm{eff,i}\big)^2 = \frac{l_x l_y l_z}{V_\mathrm{cell}} \langle H_\mathrm{eff}^2 \sin^2 \theta \rangle_\mathrm{dw}
\]
where $\langle \dots\rangle_\mathrm{dw}$ denotes the average of the value inside the domain wall, $\theta$ is the angle between the effective field and the magnetic moment, and $l_x, l_y$ and $l_z$ are the extent of the domain wall in $x$, $y$, and $z$-directions, respectively. 

When the film oscillates, the magnetization follows the external field, and the center of the domain wall prefers 
to stay at $B_\mathrm{ext}(x) = 0$. As a consequence of the speed of the magnetic relaxation, the angle 
between the field and magnetic moments $\theta$ remains small even for relatively high film velocities, 
and hence $\sin \theta \approx \theta$. The angle depends on the film velocity $v$ relative to the relaxation speed of 
the magnetization and the width of the domain wall. As such, we approximate
\[
\theta \approx c\frac{v}{\gamma_0 H_\mathrm{eff} l_x},
\]
where $c$ is a dimensionless coefficient. Inserting the approximation into (\ref{EQDissipation}) and expressing the damping coefficient as a function of the dissipated power and velocity, $\Gamma = P/v^2$, the time-dependent 
terms cancel out and we have 
\[
\Gamma \approx \frac{\alpha}{1+\alpha^2}\frac{c^2\mu_0 M_\mathrm{sat} l_y l_z}{\gamma_0 l_x}.
\]
Furthermore, using \mbox{$l_x = \pi\sqrt{A_\mathrm{ex}/K_\mathrm{u}}$} for the domain wall width, we obtain a 
ballpark estimate for the damping coefficient that depends on the four micromagnetic material parameters and the 
size of cross section of the domain wall:
\begin{equation}
\Gamma \approx \frac{\alpha}{1+\alpha^2}\sqrt{\frac{K_\mathrm{u}}{A_\mathrm{ex}}}\frac{c^2\mu_0 M_\mathrm{sat} l_y l_z}{\pi\gamma_0}.
\label{EQtheory}
\end{equation}
Using the default parameters, the size of the film of the first setup and $c=1$, the estimated damping coefficient is 
approximately $2.5\cdot10^{-14}$ kg/s. This expression is independent of time and hence is in line with the expectation of a frequency-independent $\Gamma$ in the limit of low frequencies.

\section{Results}

\begin{figure}[t!]
\begin{center}
\includegraphics[trim=0cm 0cm 0cm 0cm, clip=true,width=0.94\columnwidth]{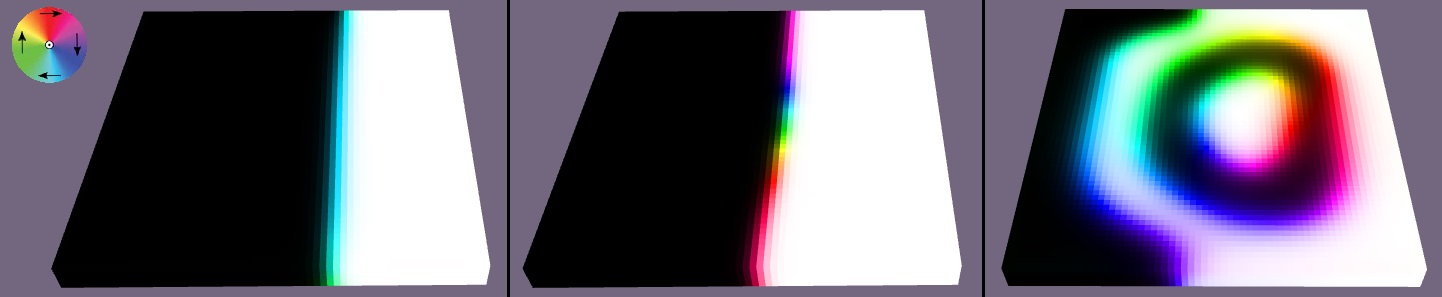}
\caption{The magnetization behavior during oscillation with frequency $f_0 = 112$ MHz in parameter regions \textbf{I} (left), \textbf{II} (center), \mbox{\textbf{III} (right)}, with $M_\mathrm{sat}$ values 350 kA/m, 170 kA/m and \mbox{510 kA/m}, respectively, with the other parameters assuming their default values. Compared to the smooth domain wall motion regime \mbox{(region \textbf{I})} where $\Gamma$ is largely independent of frequency, low $M_\mathrm{sat}$ (region \textbf{II}) cause Bloch line excitations in the inner structure of the domain wall during oscillation, strengthening the dissipation and the frequency dependence of the damping coefficient. In region \textbf{III}, the domain wall structure breaks down due to shape anisotropy and the damping is reduced.}
\label{FIGTwocase}
\end{center}
\end{figure}

In both the single thin film and the tip-strip configurations, we observed three material parameter ranges in which the systems' magnetic response to the oscillation changes significantly, with accompanying changes in the damping coefficient. Example snapshots of the magnetic configuration in the single film during oscillation in the different regimes is shown in figure~\ref{FIGTwocase}. 

In the first parameter range (indicated as region \textbf{I} in the figures of this section), the domain wall(s) in the film or substrate oscillate without internal excitations for both $f_0 = 56$ MHz and $f_0 = 112$ MHz. In this parameter regime, the change in damping coefficient when changing material parameters is relatively modest and smooth, and the damping coefficients were roughly of the same order of magnitude as the estimate \mbox{of (\ref{EQtheory}).}\vspace{-0mm}

The second parameter regime (region \textbf{II}) was characterized by excitations in the domain wall(s) of the system with the $f_0 =  112$ MHz and, depending on the parameters, possibly also with the lower frequency. For example, in the single film setup, \mbox{Bloch lines \cite{toukoBlochLine}} appeared in the domain wall during motion in simulations with low $M_\mathrm{sat}$, enhancing the dissipation and resulting in larger damping coefficients. This behavior likely results from the external field affecting the domain wall exceeding Walker field $H_\mathrm{W} = 2\pi\alpha M_\mathrm{sat}$ \cite{Walker}. In the tip-strip configuration, the domain walls exhibit precessional motion in region \textbf{II} with \mbox{$f_0 = 112$ MHz}, though without Bloch line excitations, likely due to the restricted strip width in the $y$-direction. With $f_0 = 56$ MHz the precessional motion was mostly absent.

Due to the excitations, the oscillation frequency in the second parameter regime has a strong effect on the damping, as faster velocities result in less time for the magnetization to relax and the excitations to abate. The dissipation also ceases to be strictly exponential, or splits into two different phases with different decay times for the oscillation amplitude, having an initial part with the excitations and the latter part where the excitations have died out. In these cases we fit the exponential $A_0e^{-t/\tau}$ and calculate the damping coefficients from the part with excitations and
 hence strong dissipation. 

Third parameter regime \mbox{(region \textbf{III})} consists of values with which the system turns in-plane or assumes a complex magnetic configuration due to shape anisotropy overcoming the magnetocrystalline anisotropy, causing the magnetization to twist in order to avoid stray fields. In these cases, the change in magnetization in response to motion is typically weak, and thus the damping coefficient is very low or even zero.

\subsection{Thin film in an external field}

\begin{figure}
\includegraphics[trim=2.0cm 0cm 1.5cm 0.0cm, clip=true,width=1.0\columnwidth]{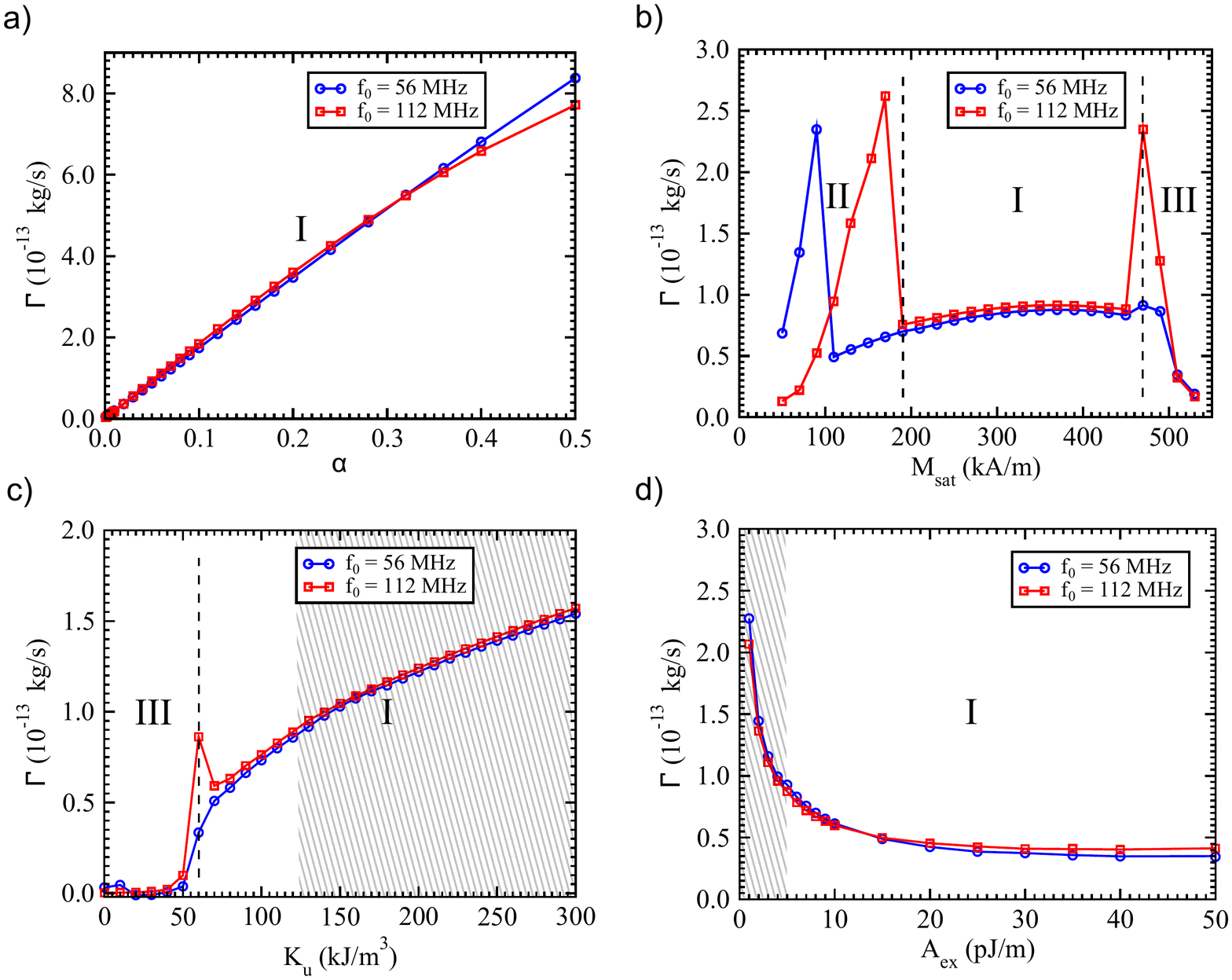}
\caption{The damping coefficient $\Gamma$ as a function of each of the micromagnetic parameters: \textbf{a)} Gilbert damping constant $\alpha$, \mbox{\textbf{b)} saturation} magnetization $M_\mathrm{sat}$, \textbf{c)} uniaxial anisotropy constant $K_\mathrm{u}$ and \textbf{d)} exchange constant $A_\mathrm{ex}$. In the no-excitation regime \mbox{(region \textbf{I})}, the film and the domain wall oscillate smoothly, and the parameter dependence of the damping coefficient is generally well-behaved.  In the parameter regime in which the inner structure of the domain wall can be excited (region \textbf{II}), the energy dissipation and the damping coefficient are substantially enhanced with the higher frequency. The parameter range where the system becomes qualitatively different (such as becoming completely in-plane polarized) typically have small or zero damping coefficients (region \textbf{III}). The shaded area shows the parameter ranges for which the smaller cell size was used.}
\label{FIGParamsFilm}
\end{figure}

In the single thin film simulations, the domain wall moves with the film, but is pushed back towards the center through the external field, the strength of which grows linearly with the distance from the center. If the mobility of the domain wall is high enough (or the oscillation slow enough), the domain wall remains located at the center without abrupt changes in the inner structure, leading to smooth magnetic moment relaxation. This was the case for most of the tested parameters, as evidenced by both frequencies exhibiting the same damping coefficient for a large range of parameters (figure~\ref{FIGParamsFilm}).

The strength of magnetic friction increases with a larger Gilbert damping constant, and in the parameter \mbox{range $\alpha = 0.001$ -- 0.50} \mbox{(figure~\ref{FIGParamsFilm} \textbf{a})}, the increase in damping coefficient is linear for the lower oscillation frequency and almost linear for the higher frequency. Since the magnetic structure itself is not influenced by $\alpha$, the close-to-linear damping coefficient dependence $\Gamma \propto\alpha/(1+\alpha^2)$ predicted by (\ref{EQtheory}) matches quite well in this parameter range. The higher values limit the velocity of the domain wall, causing it to lag behind the film in oscillation. This likely limited the dissipation and thus lowered the damping coefficient with \mbox{$112$ MHz} frequency compared to \mbox{$56$ MHz} for the highest values of $\alpha$.

The saturation magnetization $M_\mathrm{sat}$ affects the damping coefficient in a more complicated fashion, due to it influencing all the effective field terms except for the external field \cite{numericalmicromagnetics}. Additionally, the Walker breakdown threshold is lowered on small $M_\mathrm{sat}$, and as we see in \mbox{figure~\ref{FIGParamsFilm} \textbf{b}} the damping coefficient has a sharp increase for low $M_\mathrm{sat}$ values due to the domain wall excitation. The effect of the oscillation frequency can be seen as the shortened no-excitation regime at the lower values of $M_\mathrm{sat}$ \mbox{for $f_0 = 56$ MHz}, due to domain wall having more time to relax back to the center and the local field not exceeding the Walker field. Lowering $M_\mathrm{sat}$ further eventually leads to vanishing magnetization and thus to the elimination of magnetic friction.

Increasing $M_\mathrm{sat}$ increases the demagnetizing field energy greatly as compared to the exchange and magnetocrystalline anisotropy energies and thus pushes the film towards an in-plane configuration to minimize stray fields. Between \mbox{$170$ kA/m} and \mbox{$450$ kA/m}, the increased demagnetizing field energy causes the domain wall to widen and bend somewhat at the edges of the film, though the wall still oscillates smoothly without excitations, leading to only a small increase in the damping coefficient with the higher frequency. The widening of the domain wall decreases the dissipation power due to smoother change in magnetization. Though \mbox{(\ref{EQtheory})} would suggest linear increase in the damping coefficient, it is almost level in this region with both frequencies, likely due to the domain wall widening and the contributions of the other field terms canceling each other out. 

\mbox{For $M_\mathrm{sat} = 470 $ -- 500 kA/m}, the shape anisotropy starts to overcome the magnetocrystalline anisotropy, causing more prominent bending and twisting of the domain wall during oscillation. With low frequency, the effect of these fluctuations of the domain wall shape are mostly negligible, but with the high frequency the damping coefficient increases drastically, since the domain wall doesn't have adequate time to relax. For even larger $M_\mathrm{sat}$ values, the in-plane tilt changes the equilibrium magnetization to a bubble domain instead of a domain wall. The magnetization of the bubble changes relatively little during the oscillation, independently of the frequency, bringing the damping coefficient down again.

Due to their effects on the domain wall width $l_x$, changes in $A_\mathrm{ex}$ \mbox{(figure~\ref{FIGParamsFilm} \textbf{c})} and $K_\mathrm{u}$ \mbox{(figure~\ref{FIGParamsFilm} \textbf{d})} have opposite effects on the observed magnetic friction, though the domain structure at extreme values of the two parameters is different. Above $l_x\approx 3\cdot10^{-8}$ m the system quickly collapses to either a vortex ($K_\mathrm{u} < 50$ kJ/m$^3$) due to domination of shape anisotropy, or a single-domain state (\mbox{$A_\mathrm{ex} >$ 50} J/m, not shown in the figure) because of the strength of the exchange interaction combined with the strong magnetocrystalline anisotropy force the magnetic moments to the same direction. Both states have greatly reduced the damping coefficients due to negligible magnetization change during motion. 

With $l_x < 3\cdot10^{-8}$ m, the magnetic structure of two domains with a domain wall in the center is reobtained. In this regime, the domain wall oscillates smoothly, and the damping coefficient is approximately inversely proportional to the wall width $l_x$, as predicted by (\ref{EQtheory}). The noticeable spike in damping coefficient at $K_\mathrm{u} = 60$ kJ/m$^3$ with $f_0 = 112$ MHz is due to the system being between in-plane and out-of-plane polarization with this value, resulting in a greater freedom for the direction \mbox{of $\mathbf{m}$}. The less constrained spins combined with a fast oscillation frequency prevent the system from relaxing to a stable and smoothly changing configuration, resulting in large domain wall fluctuations. With $f_0 = 56$ MHz, the change of magnetization is smoother, suppressing the fluctuations.

With a high $K_\mathrm{u}$ or low $A_\mathrm{ex}$ the domain wall can be artificially pinned due to the simulation cells being too large compared to the domain wall width. The pinning and depinning during motion causes excitations in the domain wall, resulting in a rapid and strong (up to an order of magnitude) increase in damping coefficient. To avoid the artificial excitations, we switched to the smaller cell size in these cases. The shaded area in \mbox{figures~\ref{FIGParamsFilm} \textbf{c} and \ref{FIGParamsFilm} \textbf{d}} indicates the values for which the smaller cell size was used.

\subsection{Tip-strip configuration}

\begin{figure}
\begin{center}
\includegraphics[trim=0cm 0cm 0cm 0cm, clip=true,width=0.95\columnwidth]{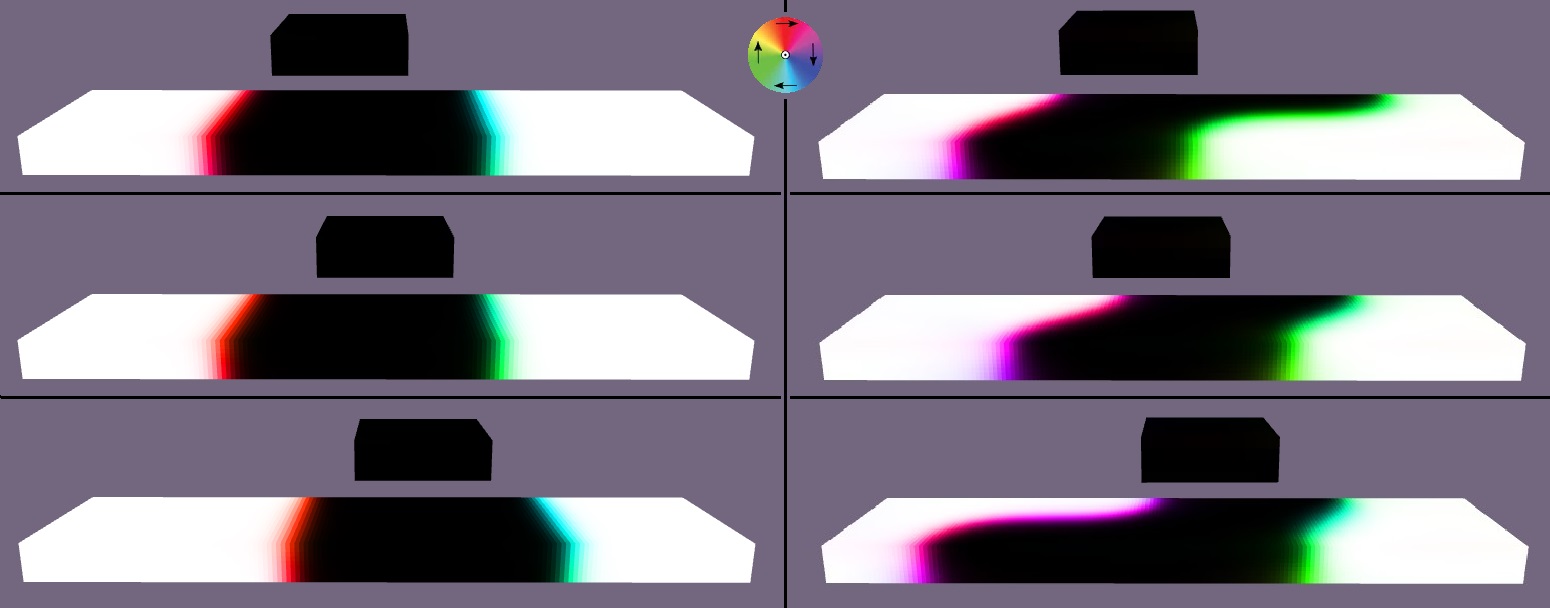}
\caption{The oscillation of tip and magnetization in the tip-strip system with \mbox{$M_\mathrm{sat} = 300$ kA/m (left)} and with $M_\mathrm{sat} = 490$ kA/m (right). The frequency is \mbox{$f_0 = 56$ MHz} and the other parameters have their default values. With high saturation magnetization the domain walls in the strip bend, and the tip motion can cause them to swiftly relax into a new configuration, resulting in fluctuations and a considerable increase in energy dissipation.}
\label{FIGBending}
\end{center}
\end{figure}

\begin{figure}
\includegraphics[trim=2.1cm 0cm 1.60cm 0.0cm, clip=true,width=1.0\columnwidth]{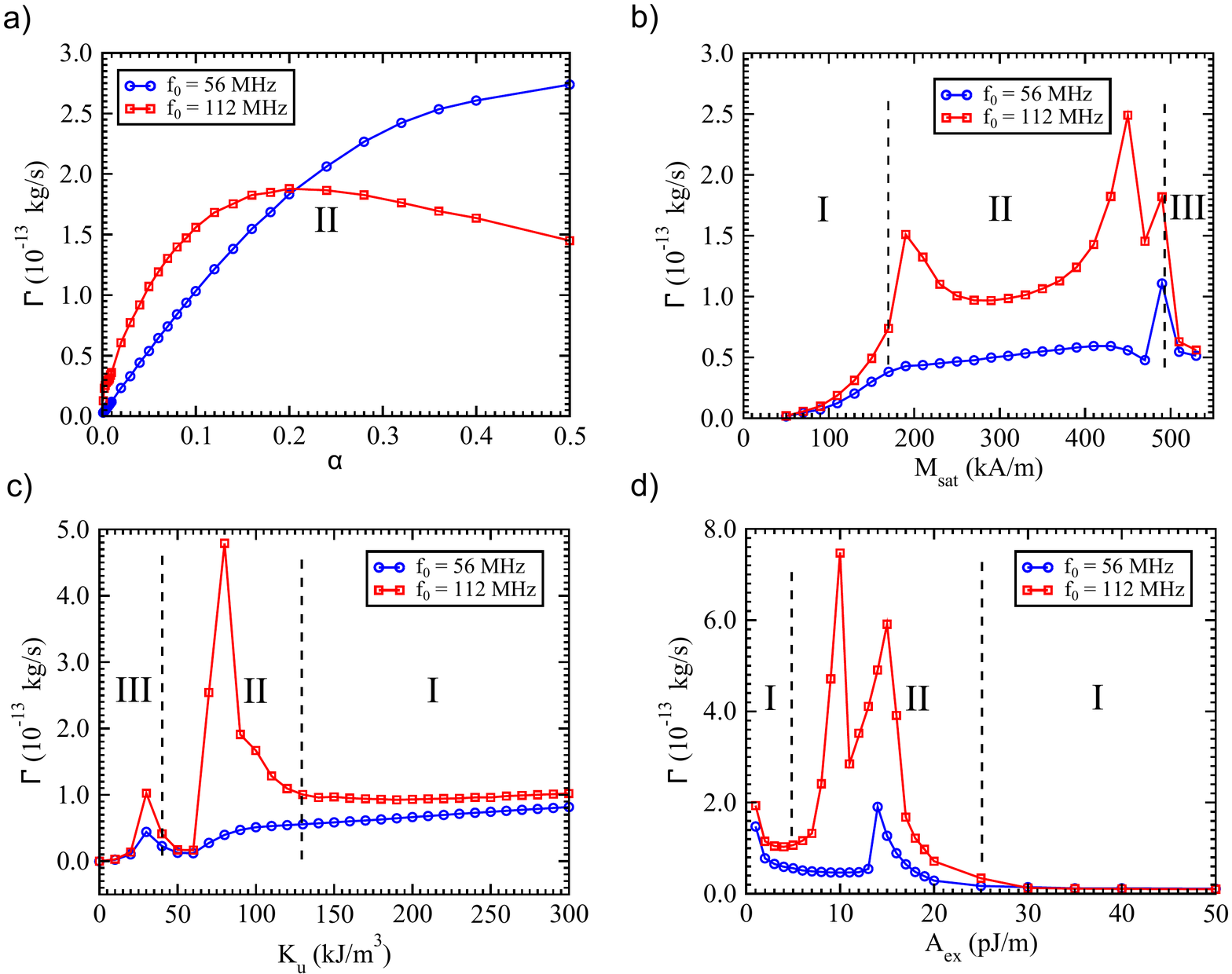}
\caption{The damping coefficient $\Gamma$ as a function of each of the micromagnetic parameters in the tip-strip configuration, laid out similarly as in figure~\ref{FIGParamsFilm}. Overall, the damping coefficient dependency on the oscillation frequency is more pronounced (especially notable in \textbf{c} and \textbf{d}) due to the precessional motion and fluctuations of the domain walls with $f_0 = 112$ MHz. The precessional motion was absent for $f_0 = 56$ MHz except at $A_\mathrm{ex} \approx 15$ pJ/m. In this setup, there are no excitations due to low $M_\mathrm{sat}$ (shown in \textbf{b}), since the domain wall movement is subdued because of the weak demagnetizing field.}
\label{FIGTipStrip}
\end{figure}

The behavior of the tip-strip system with two domain walls differs considerably from single film system. Within this configuration, the oscillation frequency has a stronger effect on the observed damping coefficient, since $f_0 = 112$ MHz caused the domain walls to exhibit precessional motion with most material parameters. The parameters also had a stronger influence to the domain wall behavior compared to the single film system, with some parameters introducing fluctuations to the system which greatly increased the damping. An example of domain wall bending and fluctuation with a particular parameter value of $M_\mathrm{sat} = 490$ kA/m, compared to a non-fluctuating case, is shown in figure~\ref{FIGBending}. In the regime with domain wall fluctuations, the dissipation wasn't always strictly exponential, and thus the decay times and the calculated damping coefficients aren't as accurate. The damping coefficient as a function of each material parameter is shown in figure~\ref{FIGTipStrip}.

Like the single thin film case, there are no internal excitations in the walls for any $\alpha$ value, though with $f_0 = 112$ MHz the domain walls in the substrate display precessional motion. The lower oscillation frequency shows a similar linear part in damping coefficient with low $\alpha$, but contrary to the single film, the increase starts evening out for high values \mbox{(figure~\ref{FIGTipStrip} \textbf{a})}. This happens because the domain walls can't keep up with the tip oscillation due to the weaker field in this configuration (compared to the external field in the single thin film case) and higher $\alpha$ reducing domain wall mobility. The domain walls lagging behind the tip leads to overall smaller domain wall movements and weaker dissipation. Using the higher oscillation frequency, the effect is further exacerbated, with $\Gamma$ reaching its peak at relatively low values, actually decreasing for higher values. In an experimental setup with lower oscillation frequencies, it's likely that the dependence of the damping coefficient is closer to linear like in the single film case, as the domain walls wouldn't lag behind the tip.

As for the saturation magnetization, we see from \mbox{figure~\ref{FIGTipStrip} \textbf{b}} that the behavior is mostly similar to the single film case, though the precessional domain wall motion with the higher oscillation frequency creates a large gap between the observed damping coefficients. For low values (\mbox{$M_\mathrm{sat} < 150$ kA/m}), the weak demagnetizing field of the tip is barely enough to move the domain walls, and as a result $\Gamma$ is low. Larger $M_\mathrm{sat}$ strengthens the stray field, making the domain walls of the substrate move more in accordance with the tip and increasing the damping coefficient. As with the single film system, \mbox{at $M_\mathrm{sat} > 450$ kA/m} the domain walls begin to twist and bend in the substrate, increasing dissipation. The spike at $\mbox{$M_\mathrm{sat} = 490$ kA/m}$ is the result of rapid reconfigurations in response to the tip motion, as was shown in figure~\ref{FIGBending}. Higher values turn the magnetizations of both films further in-plane due to shape anisotropy starting to dominate over the magnetocrystalline anisotropy, resulting a weakened demagnetizing field and hence a decrease in energy dissipation.

The effects of anisotropy constant and exchange constant with $f_0 = 56$ MHz are similar to the single film case in most respects. However, with $f_0 = 112$ MHz at the intermediate value ranges for $K_\mathrm{u}$ and $A_\mathrm{ex}$, i.e. region \textbf{II} in \mbox{figures~\ref{FIGTipStrip} \textbf{c} and \textbf{d}}, the precessional motion of the domain walls in the substrate tends to grow in intensity. Near $K_\mathrm{u} = 80$ kJ/m$^3$, the situation is quite identical to the $M_\mathrm{sat} = 490$ kA/m, i.e. the lowered magnetocrystalline anisotropy energy competes with the stray field energy, resulting in twisting of the domain walls and strong fluctuations. The exchange constant also has maxima at values around \mbox{$A_\mathrm{ex}$ = 10 pJ/m} and \mbox{$A_\mathrm{ex}$ = 15 pJ/m}, where the domain walls fluctuate heavily in response to the tip motion (figure~\ref{FIGTipStrip} \textbf{d}). To an extent, this happens also with the lower oscillation frequency, though the dissipation peak is much smaller and only occurs at $A_\mathrm{ex}$ = 14 pJ/m. The exact reason for the strong fluctuations at these particular values is unclear. Outside the intermediate value ranges, the domain wall motion is mostly smooth and the damping coefficient regains the approximate inverse dependence to the domain wall width. 

Like in the single film scenario, high values for $A_\mathrm{ex}$ and low values for $K_\mathrm{u}$ drive the system toward an in-plane configuration. For $A_\mathrm{ex} > 50$ pJ/m, the strong exchange interaction again pushes the thin film toward a single-domain configuration, resulting in the two domain walls in the substrate annihilating at the beginning of the oscillation and therefore vanishing friction. A similar thing happens with the lowest $K_\mathrm{u}$ values. However, contrary to the single film case, the in-plane configuration has a small dissipation peak also at low $K_\mathrm{u}$. In this case, the magnetization in the substrate turns in-plane with a single Néel wall. During the oscillation the wall passes under the tip, flipping its magnetization. This causes magnetic oscillations in both the tip and the substrate, dissipating energy and increasing the damping coefficient. In the rest of the in-plane configurations, magnetic friction is negligible. A further increase in $K_\mathrm{u}$ or decrease in $A_\mathrm{ex}$ will eventually lead to an unphysical situation where the domain walls cannot be resolved properly due to discretization cell size, and the magnetic friction vanishes.

\subsection{Frequency dependence of $\Gamma$}

\begin{figure}
\begin{center}
\includegraphics[trim=1cm 4cm 0.5cm 0cm, clip=true,width=1.00\columnwidth]{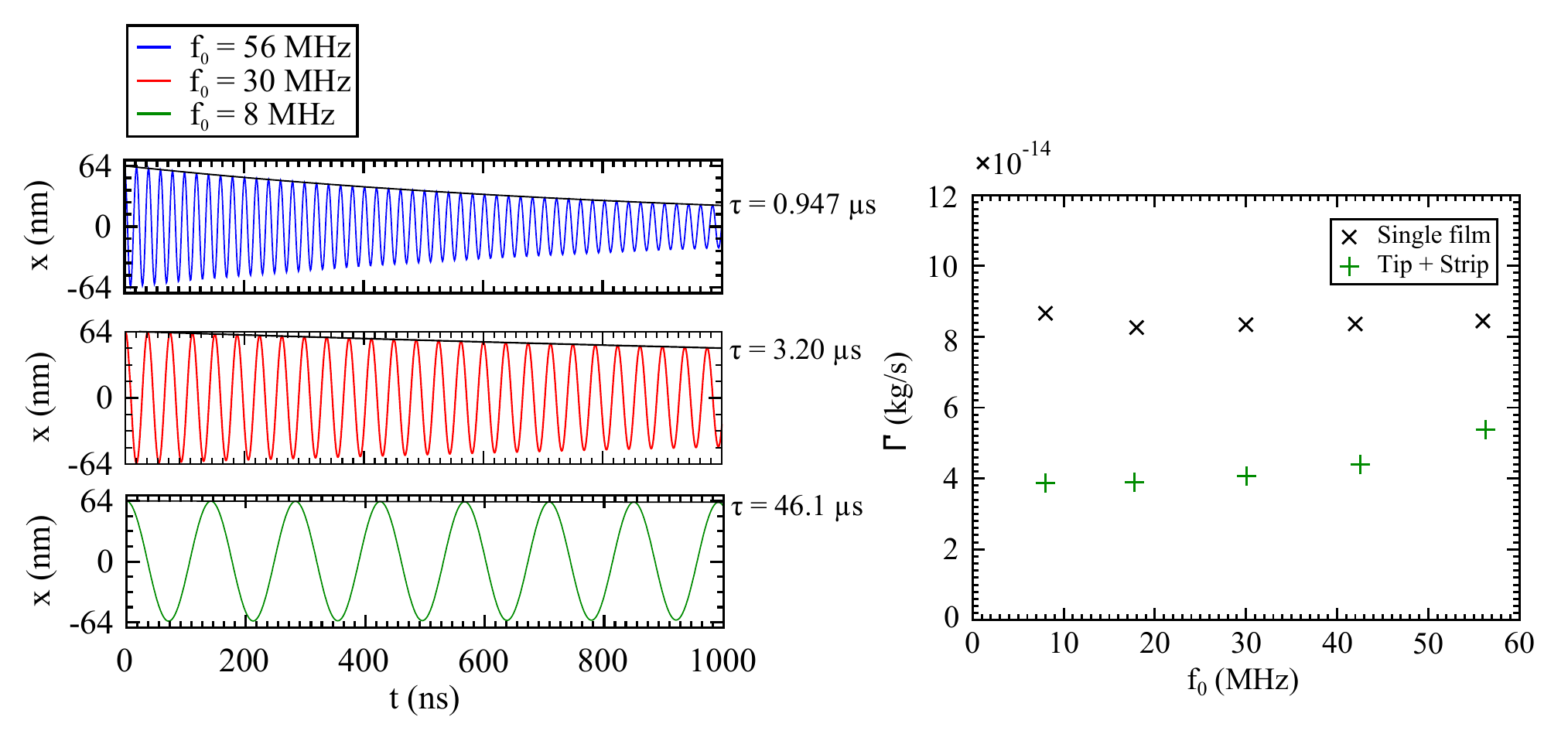}
\caption{The time evolution of the amplitude of the oscillation for three different frequencies in the single thin film setup, with the decay time constant $\tau$ indicated for each frequency (left). The damping coefficients $\Gamma$ for both simulation setups are found to be well within the same order of magnitude (right).}
\label{FIGtaus}
\end{center}
\end{figure}

As seen in the previous results, the damping coefficient $\Gamma$ can depend weakly or strongly on the frequency, depending on whether the domain wall oscillates smoothly or experiences fluctuations or excitations during motion. To further investigate whether the damping coefficient is affected by the frequency in the smooth domain wall motion regime, we ran simulations for both the single film and the tip-strip setups with oscillation frequencies ranging from the \mbox{$f_0 = 56 $ MHz} down to 
\mbox{$f_0 = 8$ MHz}. In these simulations we use the default parameters, which for both setups are in region I, meaning that the domain walls move without excitations or fluctuations already for \mbox{$f_0 = 56 $ MHz}.

Though the oscillation did not have enough time to fully die down with the lower frequencies considered, fitting the exponential function to the decaying amplitude we find that the damping coefficients are close in magnitude, with the tip+strip system showing only a slight increase towards $f_0 = 56$ MHz and the single film setup not displaying any clear trend (Fig.~\ref{FIGtaus}). This result suggests that for domain wall motion without excitations, found at frequencies well below the Larmor precession frequency of the magnetic moments, the damping coefficient $\Gamma$ is independent of the mechanical oscillation frequency. 

The frequency independence is also evident from Brown's expression for energy dissipation, equation (\ref{EQBrown}). The magnetization only changes inside the domain walls, and for low frequency oscillation the magnetization of the domain wall adapts to the change of position of the oscillating film practically instantly, meaning that the domain wall moves at the same velocity $v$ as the film and can be considered to move rigidly. It follows that the time derivative of the domain wall magnetization is directly proportional to the velocity, $d\mathbf{m}/dt \propto v$, and from (\ref{EQBrown}) we see that $P \propto v^2$. Since $\Gamma = P/v^2$, the damping coefficient becomes independent of the oscillation frequency. These results are in line with the notion that excess (or anomalous) losses in magnets with bar-like domains, arising from domain wall motion, dissipate power as $P \propto v^2$ \cite{Bishop_1976}. 

\section{Conclusion}

We have simulated ring-down measurements using micromagnetics, investigating how domain wall motion dissipates energy and causes damping of mechanical oscillation of magnetic thin films. Our results indicate a rather complex relationship between oscillation frequency, material parameters and energy dissipation. For low frequencies the material parameters have a relatively weak influence on magnetic friction, whereas high frequencies can have a dramatic effect due to domain wall excitations. In the latter case, magnetic friction was found to be especially strong in situations where no energy term strongly dominated the system and the domain wall shape fluctuated during motion. The observed damping coefficients were of roughly the same order of magnitude as those found in experiments concerning other forms of non-contact friction. The damping coefficients are also in the same range as those found in the experiment of \cite{tipexperiment} for the cobalt tip oscillating in an external field.

Compared to typical cantilever systems, the mechanical oscillation frequencies used in the simulations are considerably higher, as simulating closer to experimental frequencies with micromagnetics is challenging due to the picosecond timescale of magnetization dynamics compared to the relaxation time of the oscillation. However, as various excitations are already weak to nonexistent at the 56 MHz frequency, it is unlikely that the still lower frequency of real systems would introduce any new dynamics to the system that would significantly alter the damping coefficient. This is especially true in parameter regime where the motion of the domain wall is smooth and without excitations, where lowering frequency down to 8 MHz did not affect the damping coefficient for either of the simulated systems.

The simulations of this work show that interesting dynamics emerge when mechanical motion and magnetic domain evolution are coupled. Larger systems with disorder and more complex magnetic structure will remain as a prospect for future study.

\ack
\noindent We acknowledge the support of the Academy of 
Finland via an Academy Research Fellowship (LL, projects no. 268302 and 303749), and 
the Centres of Excellence Programme (2012-2017, project no. 251748). We acknowledge 
the computational resources provided by the Aalto University School of Science 
Science-IT project and the CSC. 

\section*{References}
\bibliographystyle{iopart-num}
\bibliography{bibl}

\end{document}